\documentclass[apj]{emulateapj}
\usepackage{graphics}
\usepackage{epsfig}

\newcommand{\oiiiw}{\mbox{[\ion{O}{3}] $\lambda$5007} $\,$}
\newcommand{\oiiiwn}{\mbox{[\ion{O}{3}] $\lambda$5007}}
\newcommand{\hb}{\mbox{H$\beta$} $\,$}
\newcommand{\hbn}{\mbox{H$\beta$}}
\newcommand{\ha}{\mbox{H$\alpha$} $\,$}
\newcommand{\han}{\mbox{H$\alpha$}}
\newcommand{\oiiihb}{\mbox{[\ion{O}{3}] $\lambda$5007}/{\mbox{H$\beta$} $\,$}}
\newcommand{\oiiihbn}{\mbox{[\ion{O}{3}] $\lambda$5007}/{\mbox{H$\beta$}}}

\newcommand{\nii}{\mbox{[\ion{N}{2}] $\lambda$6584} $\,$}
\newcommand{\niin}{\mbox{[\ion{N}{2}] $\lambda$6584}}
\newcommand{\niiha}{\mbox{[\ion{N}{2}] $\lambda$6584}/{\mbox{H$\alpha$} $\,$}}
\newcommand{\niihan}{\mbox{[\ion{N}{2}] $\lambda$6584}/{\mbox{H$\alpha$}}}

\slugcomment{\it Submitted for Publication in ApJL}
\shortauthors{Comerford et al.}
\shorttitle{A 1.75 $h^{-1}$ kpc Separation Dual AGN at $z=0.36$ in the COSMOS Field}
\begin{document}

\title{A 1.75 $\MakeLowercase{h^{-1}}$ kpc Separation Dual AGN at $\MakeLowercase{z}=0.36$ in
  the COSMOS Field}

\author{Julia M. Comerford\altaffilmark{1}, Roger L. Griffith\altaffilmark{2}, Brian
  F. Gerke\altaffilmark{3}, Michael C. Cooper\altaffilmark{4,5}, \\ Jeffrey A. Newman\altaffilmark{6}, Marc
Davis\altaffilmark{1,7}, and Daniel Stern\altaffilmark{2}}

\affil{$^1$Astronomy Department, 601 Campbell Hall, University of
California, Berkeley, CA 94720}
\affil{$^2$Jet Propulsion Laboratory, California Institute of Technology, MS 169-327, 4800 Oak Grove Drive, Pasadena, CA 91109}
\affil{$^3$Kavli Institute for Particle Astrophysics and Cosmology,
  M/S 29, Stanford Linear Accelerator Center, \\ 2575 Sand Hill Rd., 
  Menlo Park, CA 94725}
\affil{$^4$Steward Observatory, University of Arizona, Tucson, AZ 85721}
\affil{$^5$Spitzer fellow}
\affil{$^6$Department of Physics and Astronomy, University of
  Pittsburgh, Pittsburgh, PA 15260}
\affil{$^7$Department of Physics, University of California, Berkeley, CA 94720} 

\begin{abstract}
We present strong evidence for dual active galactic nuclei (AGN) in the $z=0.36$ galaxy COSMOS
J100043.15+020637.2.  
COSMOS {\it Hubble Space Telescope} ({\it HST}) imaging of the galaxy
shows a tidal tail, indicating that the galaxy recently
underwent a merger, as well as
two bright point sources near the galaxy's center. Both the
luminosities of these sources (derived from the {\it HST} image) and
their emission line flux 
ratios (derived from Keck/DEIMOS slit spectroscopy) 
suggest that both are AGN and not star-forming
regions or supernovae.  Observations from zCOSMOS, Sloan Digital Sky Survey, {\it XMM-Newton}, Very
Large Array, and {\it Spitzer} 
fortify the evidence for AGN activity.  With {\it HST} imaging we
measure a projected spatial offset between the two 
AGN of 1.75 $\pm$ 0.03 $h^{-1}$ kpc, and with DEIMOS we measure a
150 $\pm$ 40 km s$^{-1}$ line-of-sight velocity offset 
between the two AGN. Combined, these observations provide
substantial evidence that COSMOS
J100043.15+020637.2 is a dual AGN in a merger-remnant galaxy. 
\end{abstract}


\section{Introduction}
\label{intro}
In the standard $\Lambda$ cold dark matter paradigm of structure
formation, more massive galaxies are assembled from smaller ones in a
series of merger events.  Nearly every galaxy hosts a central
supermassive black hole (SMBH) \citep{KO95.1}, which implies that a
merger between two galaxies nearly always results in a
merger-remnant galaxy containing two SMBHs.  Drag from dynamical
friction causes the two SMBHs to inspiral toward the center of the
merger-remnant. The SMBHs spend $\sim 100$ Myr at separations $\gtrsim
1$ kpc \citep{BE80.1, MI01.1}, then form a parsec-scale binary and
ultimately coalesce into a single central SMBH in the merger-remnant
galaxy.  This final coalescence is necessary to preserve the tight
observational correlation between the mass of the black hole and the
velocity dispersion, or total mass, of the host galaxy stellar bulge \citep{FE00.1}.  

Although SMBH pairs are a natural consequence of galaxy mergers, there
have been few unambiguous detections of galaxies hosting SMBH pairs.
If sufficient gas accretes onto both SMBHs, they may each be visible as an
active galactic nucleus (AGN).  To date, there have been definitive
detections of only four galaxies hosting such AGN pairs.  First, radio
signatures of AGN activity in the
$z=0.055$ elliptical galaxy 0402+379 show that it hosts binary
SMBHs separated by 5 h$^{-1}$ pc \citep{XU94.1,MA04.2,RO06.1}. In addition,
X-ray detections of a dual AGN in the $z=0.024$ ultraluminous infrared
galaxy NGC~6240 indicate it hosts two SMBHs separated by 0.5
$h^{-1}$ kpc 
\citep{KO03.1}.  Finally, optical spectroscopic signatures of dual AGN in the
red galaxies EGSD2 J142033.6+525917 at $z=0.71$ and EGSD2
J141550.8+520929 at $z=0.62$ show these galaxies host dual SMBHs at
separations of 0.84 $h^{-1}$ kpc and 1.6 $h^{-1}$ kpc, respectively
\citep{GE07.2,CO09.2}.  
A fifth possible example has been proposed by \cite{BO09.2}, though it
is likely to be an object of a different nature (e.g., \citealt{CH09.3,CH09.4,GA09.1,LA09.1,WR09.1}).

Here we present evidence for a 1.75 $\pm$ 0.03 $h^{-1}$ kpc projected spatial separation
and 150 $\pm$ 40 km s$^{-1}$ line-of-sight velocity separation SMBH
pair, visible as the $z=0.36$ dual AGN system COSMOS
J100043.15+020637.2 in the COSMOS field \citep{SC07.6}.  The candidate
was found serendipitously while visually inspecting postage stamp images
of COSMOS galaxies with high S\'{e}rsic indices
in the ACS-GC catalog (Griffith et al.,
in preparation), which includes
morphology measurements for over half a million
sources from five large {\it Hubble Space Telescope} Advanced
Camera for Surveys ({\it HST} ACS) imaging datasets.  In addition, the candidate has been previously classified as an AGN by both COSMOS \citep{GA09.3} and the Sloan
Digital Sky Survey (SDSS) \citep{YO00.2, RI02.2}. 
We assume a Hubble constant $H_0 =100 \, h$ km s$^{-1}$ Mpc$^{-1}$,
$\Omega_m=0.3$, and $\Omega_\Lambda=0.7$ throughout, and all distances
are given in physical (not comoving) units.

\section{Observations}

\begin{figure}[!t]
\vspace{-.8in}
\hspace{-.5in}
\begin{center}
\hspace{-1.15in}
\includegraphics[height=4.5in]{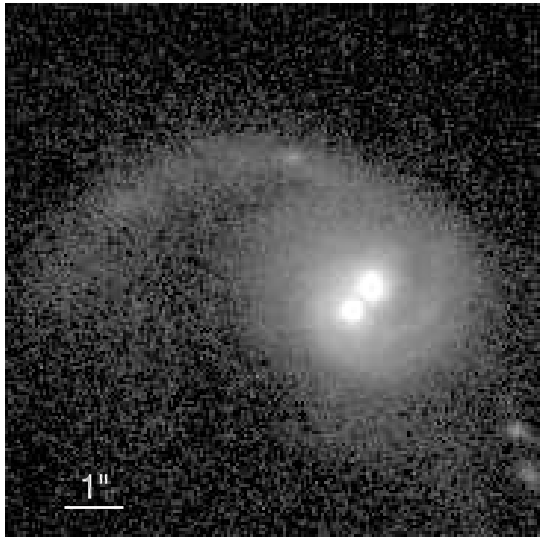}
\end{center}
\vspace{-.8in}
\caption{{\it HST} F814W ACS image of COSMOS J100043.15+020637.2,
  where North is up and East is to the left.  The galaxy's tidal tail
strongly suggests it has recently undergone a merger, and the two
bright nuclei near the
galaxy center appear to both be AGN. The nuclei are separated by $0\farcs497 \pm 0\farcs009$, or $1.75 \pm 0.03$
$h^{-1}$ kpc.}
\label{fig:cosmos}
\end{figure}

\subsection{COSMOS Imaging}
\label{cosmosim}

We originally identified COSMOS J100043.15+020637.2 as a dual AGN
candidate from its {\it HST} F814W ACS 
image taken for COSMOS \citep{SC07.7}.  This image, shown in
Figure~\ref{fig:cosmos}, shows a disturbed galaxy with a long tidal
tail that suggests the galaxy has recently undergone a merger.  The nucleus 
of the galaxy contains
two bright point sources, and we use $0\farcs25$
($0.88$ h$^{-1}$ kpc) radius apertures 
to measure the magnitude and luminosity
of each source with Source EXtractor \citep{BE96.1}.  

We measure F814W apparent magnitudes of 20.0 for the northern source and 20.1
for the southern source. For comparison, the entire object has an
apparent magnitude of 18.3, as measured using GALFIT (\citealt{PE02.4}; Griffith et al., in preparation).  The magnitudes of the point sources correspond to luminosities of 
$3.1 \times 10^{43} \; h^{-2}$ erg s$^{-1}$ ($8.0 \times 10^{9} \; h^{-2} \;
L_\odot$) for the northern source and
$2.7 \times 10^{43} \; h^{-2}$ erg s$^{-1}$ ($7.1
\times 10^{9} \; h^{-2} \; L_\odot$) for the southern source.  
Supernovae this luminous are extremely rare \citep{SM07.1},  
suggesting that the sources are most likely AGN and not supernovae.
However, the {\it HST} image was taken on UT 2004 March 16 and a more recent  
high-resolution image
would resolve this question definitively.  Regardless, other
observations strongly support the interpretation that both sources are
AGN (\S~\ref{1dspectra}).

With Source EXtractor we also find that the projected 
separation between the barycenters of the sources is
$0\farcs497 \pm 0\farcs009$, or $1.75 \pm 0.03$
$h^{-1}$ kpc, and the barycenters of the sources are aligned along a
position angle $\theta=-40^{\circ}.4$ East of North. 

\begin{figure}[!t]
\vspace{-.75in}
\hspace{-.5in}
\begin{center}
\includegraphics[width=3.5in]{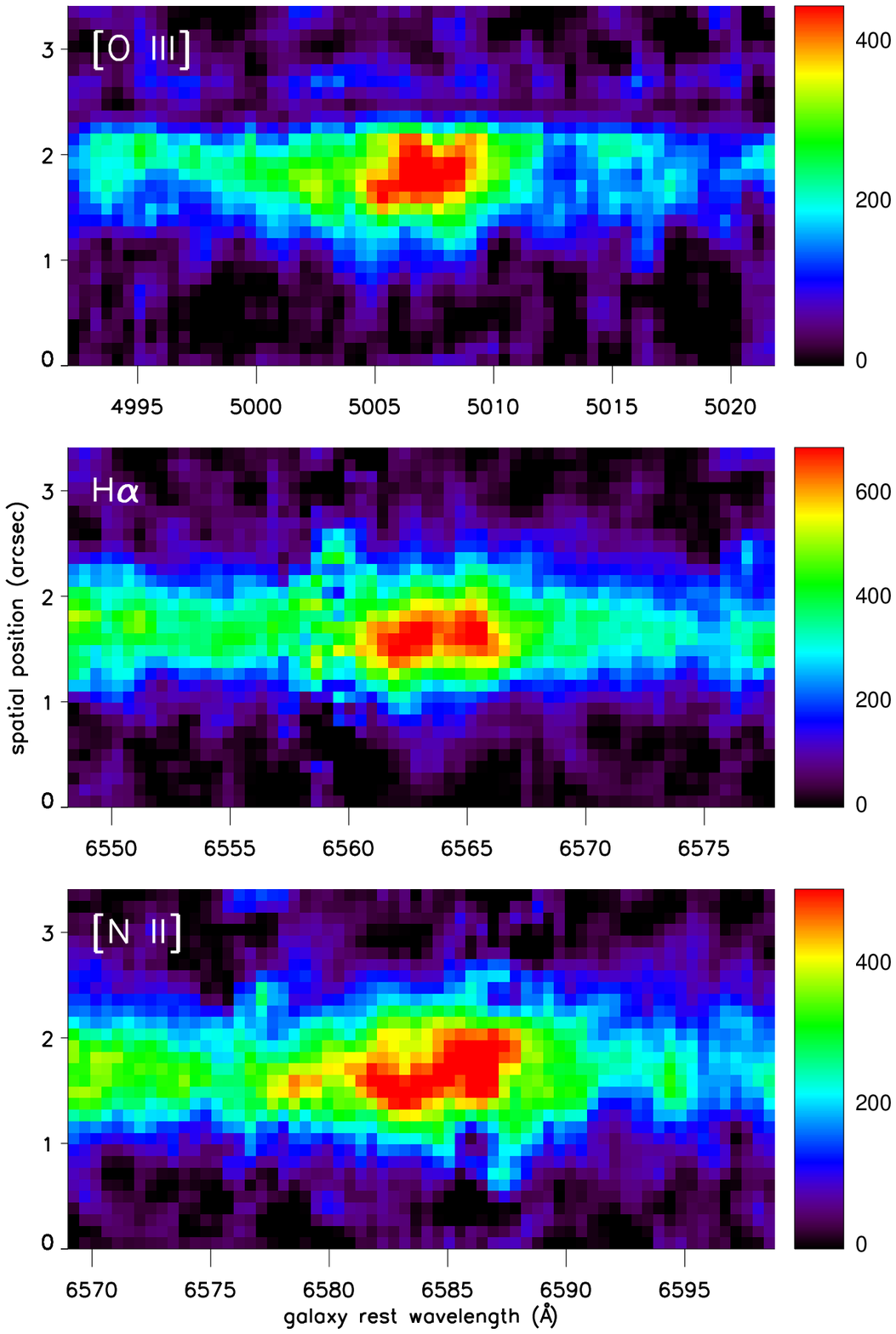}
\end{center}
\vspace{-.5in}
\caption{Two-dimensional DEIMOS spectrum of J100043.15+020637.2
  with night-sky emission features subtracted.  The spectrum has been
  smoothed by a smoothing length of 2 pixels, and AGN line emission at
  \oiiiw (top), \ha (middle), and \nii 
  (bottom) is shown.  In each panel, the vertical axis spans
  $3\farcs41$ (12.0 $h^{-1}$ kpc at the $z=0.36$ redshift of the galaxy) in spatial position along the slit 
  and the horizontal axis spans 30 \AA $\,$ in rest-frame wavelength
  centered on the emission feature. 
  The color bars provide scales for the
  flux in counts hour$^{-1}$ pixel$^{-1}$.  Each emission feature has
two components, likely corresponding to two distinct AGN, and the
velocity separation between the two is 150 $\pm$ 40 km s$^{-1}$.} 
\vspace{-.2in}
\label{fig:deimos}
\end{figure}

\begin{figure*}[!t]
\hspace{-.5in}
\begin{center}
\includegraphics[width=7.in]{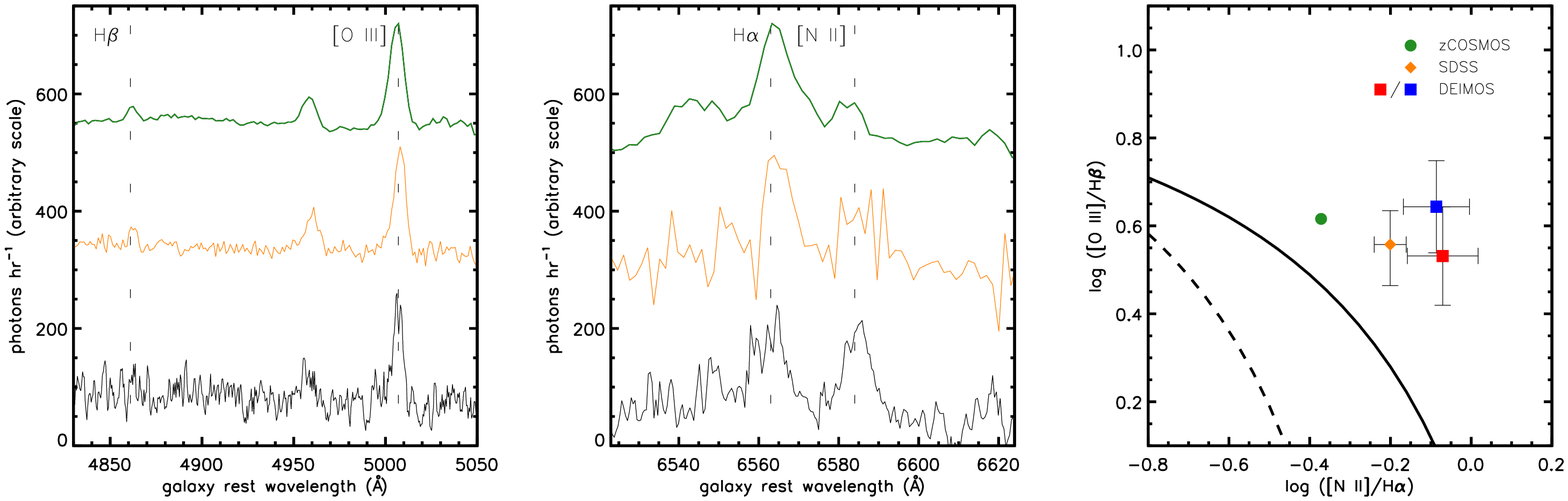}
\end{center}
\caption{Segments of the zCOSMOS (top, green), SDSS
  (middle, orange), and DEIMOS (bottom, black) spectra and the
  corresponding BPT diagram for COSMOS
  J100043.15+020637.2.  Each
  spectrum is shifted to the rest frame of the host galaxy, and the
  dashed vertical lines show the expected wavelengths of \hb and \oiiiw (left panel) and
  \ha and \nii (middle panel). For clarity, the spectra are offset from each
  other vertically and normalized to span 300 counts hr$^{-1}$. The plotted
  DEIMOS spectrum has been smoothed by taking the
  inverse-variance-weighted mean in a rolling 4 \AA $\,$ wide window.  The
  right panel shows the BPT diagram of the 
  line flux ratios for zCOSMOS (green circle), SDSS (orange diamond), and DEIMOS
  redward (red square) and blueward (blue square) emission
  components. The dashed curve illustrates the theoretical
  maximum for starbursts, based on the upper limit of stellar
  photoionization models \citep{KE01.2}, while the solid curve shows
  the empirical division between galaxies that are purely star-forming
  and ``composite'' galaxies whose spectra are dominated by
  both star formation and 
  AGN \citep{KA03.1}.  Pure star-forming galaxies lie under the dashed
  curve, composite galaxies lie between the dashed and solid curves,
  and pure AGN galaxies lie above the solid curve.  Our line
  flux ratio measurements of COSMOS
  J100043.15+020637.2 show that both of its emission components are
  dominated exclusively by AGN activity. }
\label{fig:spectra}
\end{figure*}

\subsection{DEIMOS Slit Spectroscopy}
\label{deimos}
We used the DEIMOS spectrograph on the
Keck II telescope to obtain a 200 s spectrum of the object with a 600
lines mm$^{-1}$ grating at twilight on UT 2009 April 23.  The spectrum
spans the wavelength range 4730 -- 9840 \AA, and the position angle of the slit was $\theta=-36^{\circ}.7$ East of North.
The main purposes of the slit spectroscopy were to verify that both
central point sources were AGN and to measure the spatial and velocity
separations between the two AGN emission components.  The AGN emission is
clearly visible in \oiiiwn, \han, and \niin, as shown in
Figure~\ref{fig:deimos}. We determine the projected spatial separation for each of
these three emission features by measuring the spatial centroid of each
emission component individually.

To measure the spatial centroid of an emission component, first we center a
2 \AA $\,$ (rest-frame) wide window on the emission.  At each spatial
position we then sum the flux, weighted
by the inverse variance, over all wavelengths within the
window.  We use the spatial position where the summed flux is maximum
as the center of a 10 pixel window (where 1 DEIMOS pixel spans
$0\farcs11$) around the emission.  We then fit a quadratic to the summed
flux to locate a peak, define a narrow window centered on the peak flux,
and finally compute the line centroid within this window. We derive the error
on this spatial centroid by repeatedly adding noise to the spectrum
drawn from a Gaussian with variance matching the DEIMOS pipeline
(Newman et al., in preparation) estimate for a given pixel and redoing all
centroid measurements. 

We find the two AGN emission components
have projected separations of 1.5 $\pm$ 0.5 $h^{-1}$ kpc, 0.97 $\pm$ 0.42 $h^{-1}$
kpc, and 1.6 $\pm$ 0.4 $h^{-1}$
kpc in \oiiiwn, \han, and \niin, respectively.  
The low spatial
separation measured for the \ha emission is due to the
imperfectly-subtracted night sky line partially obscuring the blueward
portion of the \ha emission, but the \oiiiw and \ha spatial offsets
are roughly consistent with the spatial offset measured in the {\it
  HST} image (\S~\ref{cosmosim}).  We note that the 3$^{\circ}$.7 difference between the position angle of the DEIMOS
slit and the orientation of the two AGN (measured in
\S~\ref{cosmosim} from the {\it HST} image) produces a negligible (0.2$\%$)
difference between the DEIMOS and {\it HST} measurements of projected
spatial offsets.

\subsection{One-Dimensional zCOSMOS, SDSS, and DEIMOS Spectra}
\label{1dspectra}

We analyze spectra of COSMOS J100043.15+020637.2 from the
zCOSMOS spectroscopic redshift survey of the COSMOS field
\citep{LI07.3}, SDSS, and DEIMOS to both determine the galaxy's redshift and measure emission line flux ratios to determine the source of its line emission.

To measure the redshift of COSMOS J100043.15+020637.2, we determine the value
 of $z$ which minimizes $\chi^2$ when comparing to a template
 spectrum.  We
mask out all emission lines, then fit a continuum template spectrum
based on \cite{BR03.1} stellar-population synthesis models.  
The
template spectrum, described in \cite{YA06.1}, consists of a 0.3 Gyr,
solar metallicity, young stellar population combined with a
7 Gyr, solar metallicity, old stellar population.  
From the template
spectrum fits, we find that the
zCOSMOS, SDSS, and DEIMOS spectra all give consistent redshifts of $z=0.36$.

The source of line emission in a galaxy is commonly identified using
the Baldwin-Phillips-Terlevich (BPT) diagram of line ratios \citep{BA81.1,
  KE06.1}. 
To determine the source of
the line emission in COSMOS J100043.15+020637.2, we examine its \hbn,
\oiiiwn, \han, and \nii emission lines in the zCOSMOS, SDSS, and DEIMOS
spectra (Figure~\ref{fig:spectra}).

The higher spectral resolution ($R \sim 3000$) of the DEIMOS spectrum enables us to
discern substructure in the emission lines that is unresolved in the zCOSMOS and SDSS spectra ($R \sim 600$ and $R \sim
1800$, respectively).  To determine the velocity difference between
the two peaks we fit two Gaussians to the continuum-subtracted \oiiiw line profile, which 
is the emission line with 
the highest signal-to-noise ratio.  Based on the
wavelengths of the peaks of the best-fit Gaussians, we find the
line-of-sight velocity
difference between the double peaks is 150 $\pm$ 40 km s$^{-1}$.  The error in velocity is derived from
the errors in the peak wavelengths of the best-fit Gaussians added in quadrature.

To identify whether the double-peaked lines
correspond to two AGN, we examine the line flux ratios of each
emission component separately.  We fit two Gaussians to each
double-peaked line, setting the velocity separation of the peaks to the
best-fit velocity difference found for the \oiiiw line profile, but allowing
the heights and widths of the Gaussians to vary.
The areas under
the best-fit Gaussians provide estimates of the line fluxes for each
emission component, and we find \oiiihbn=$4.4 \pm 1.2$ and \niihan=$0.82
\pm 0.17$ for the blueward emission component and \oiiihbn=$3.4 \pm 1.0$ and
\niihan=$0.85 \pm 0.19$ for the redward emission component, where the
uncertainties are derived from propogation of errors in the parameters of the
best-fit Gaussians. As
Figure~\ref{fig:spectra} shows, the locations of these line flux
ratios
on the BPT diagram indicate that the line emission for both emission components is clearly produced by AGN activity and not star
formation \citep{KE06.1}.  In other words, the combined DEIMOS line flux ratios indicate that COSMOS
J100043.15+020637.2 hosts two distinct AGN at nearly the same redshifts.

To measure the line flux ratios for the lower-resolution zCOSMOS and
SDSS spectra,
we subtract the
continuum from each spectrum and then measure the flux of the \hbn,
\oiiiwn, \han, and \nii emission lines.  
For zCOSMOS we measure \oiiihbn=4.1 and \niihan=0.42 (we
cannot compute errors on these ratios because no error array is
provided with public zCOSMOS spectra; however, the error on \niiha may
be large because of the night sky line partially obscuring the \ha emission), and for SDSS we
measure \oiiihbn=$3.6 \pm 0.7$ and \niihan=$0.63 \pm 0.06$.  
The locations of these line flux
ratios
on the BPT diagram both confirm that the line emission in COSMOS
J100043.15+020637.2 is produced by AGN activity (Figure~\ref{fig:spectra}).

\subsection{Multiwavelength Detections}
\label{multiwave}

We use the Infrared Science Archive$\footnote[1]{http://irsa.ipac.caltech.edu/}$ to obtain additional observations of COSMOS J100043.15+020637.2 across different
wavebands, which bolster the evidence for AGN activity found in \S~\ref{cosmosim}
and \S~\ref{1dspectra} as described below.  

{\it XMM-Newton} detected fluxes of $3.82 \times 10^{-14} $
erg cm$^{-2}$ s$^{-1}$, $6.08 \times 10^{-14}$ erg cm$^{-2}$ s$^{-1}$,
and $2.87 \times 10^{-14}$ erg cm$^{-2}$ s$^{-1}$ at energy bands 0.5
-- 2.0 keV, 2.0 -- 10.0 keV, and 5.0 -- 10.0 keV, respectively.  These
fluxes are above the limit for AGN detection, given by $\sim 10^{-15}$
erg cm$^{-2}$ s$^{-1}$ in the 0.5 -- 2.0 keV or 2.0 -- 10.0 keV energy
bands \citep{BR07.2,CA07.2}.  In addition, the Very Large Array (VLA) detected a flux of 0.139 mJy at an observing
frequency of 1.4 GHz, which is above the limit of 0.1 mJy at 1.4 GHz for AGN detection \citep{SC07.5}.
Finally, the {\it Spitzer} Infrared Array Camera (IRAC)
measured Vega magnitudes of 14.92, 14.34,
13.79, and 12.66 at 3.6, 4.5, 5.8, and 8.0 $\mu$m, respectively \citep{SA07.1}, and
these measurements fall within the observed range of mid-infrared
colors of AGN \citep{ST05.2}.

\begin{figure}[!t]
\begin{center}
\includegraphics[angle=90,width=3.5in]{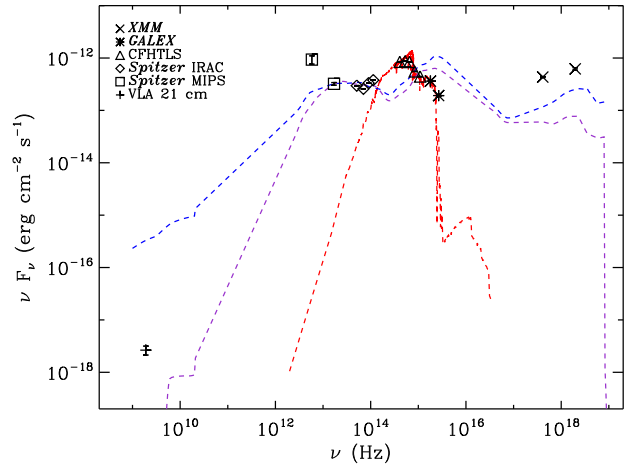}
\end{center}
\caption{SED of COSMOS J100043.15+020637.2, from radio to X-ray.  Data points
show fluxes (or upper limits) measured by various
instruments, as noted in the legend.  All frequencies have been shifted to the host galaxy's rest
frame.  Also shown are the \citet{EL94.1} median SEDs for radio-loud and radio-quiet QSOs (blue and
purple, respectively) normalized to the 24-micron flux, as well as a \citet{BR03.1} template SED for a
100-Myr-old stellar population (red) normalized to the $z'$-band flux. }
\label{fig:sed}
\end{figure}

These observations, combined with additional data from {\it Spitzer}
MIPS (at 24 and 70 $\mu$m; \citealt{SA07.1}),
the Canada-France-Hawaii Telescope Legacy Survey$\footnote[2]{http://www.cfht.hawaii.edu/Science/CFHLS/}$ (CFHTLS; in $u^{*}$,
$g'$, $r'$, $i'$, and $z'$), SDSS (in $u'$,
$g'$, $r'$, $i'$, and $z'$), and the
{\it Galaxy
Evolution Explorer} ({\it GALEX}; in the near-ultraviolet and far-ultraviolet) enable us to compile a rest-frame SED for
COSMOS J100043.15+020637.2 across several wavebands (Figure~\ref{fig:sed}).
This figure also shows the median
radio-loud and radio-quiet QSO SEDs 
compiled by \citet{EL94.1}, and a \cite{BR03.1} template SED for a
100-Myr-old simple stellar 
population with metallicity of 0.02 times the Solar value.  The QSO templates have
been normalized to the 24$\mu$m 
detection and the stellar template has been normalized to the $z'$-band
detection.  The SED of this object is broadly consistent with a
QSO that is obscured by dust in the 
optical and UV, allowing stellar emission from the host galaxy to
dominate, with strong re-radiation by 
dust in the mid-infrared.  Some or all of the infrared radiation might
also arise from ongoing 
star formation, as a merger like this one would be expected to produce
strong nuclear starbursts, but 
the X-ray and radio detections confirm the presence of at least one
active nucleus in this system. 

\section{Conclusions}

The observations discussed here provide strong evidence for a dual AGN
with 1.75 $\pm$ 0.03 $h^{-1}$ kpc projected spatial separation and 150
$\pm$ 40 km s$^{-1}$ line-of-sight velocity separation in the $z=0.36$ galaxy COSMOS
J100043.15+020637.2.  The \oiiihb and \niiha line flux ratios of the DEIMOS double-peaked emission lines are
solid indicators that the galaxy hosts two AGN, a conclusion that is bolstered by the AGN-like
luminosities of the galaxy's two central point sources.  Line flux
ratios of the unresolved, combined emission lines in zCOSMOS and SDSS spectra, as well as X-ray, radio, and infrared detections, provide further confirmation of AGN activity in the galaxy.  Finally, the tidal
tail visible in {\it HST} imaging is an unmistakable signature of a
galaxy merger.  We conclude
that COSMOS J100043.15+020637.2 is a merger-remnant galaxy with two
inspiralling supermassive black holes, each of which powers an AGN.  

The discovery of this dual AGN adds significantly to the
number of such known objects.  A statistical sample of dual AGN
would provide a direct observational probe of both the
galaxy merger rate and the kinematics of SMBH mergers, which
are expected to produce gravity waves observable by next-generation
projects such as LISA \citep{BE98.2}.

We initially identified COSMOS J100043.15+020637.2 as a dual AGN candidate
because of its two bright central sources visible in {\it HST} imaging.  Our finding was
serendipitous, and there are likely more dual AGN to be discovered in COSMOS.
We have demonstrated that dual AGN candidates can be selected as
bright double sources in {\it HST} imaging
and confirmed
through optical spectroscopy and multiwavelength observations.

\acknowledgements J.M.C. acknowledges support from NSF grant
AST-0507428.  The work of R.L.G. and D.S. was carried out at the Jet Propulsion
Laboratory at the California Institute of Technology,
under a contract with NASA. B.F.G. acknowledges support by the U.S. Department of Energy
under contract number DE-AC3-76SF00515.  M.C.C. was supported by
NASA through the Spitzer Space Telescope Fellowship program. We thank
Hai Fu for valuable assistance with the zCOSMOS spectrum.

This letter is partly based on observations carried out using
MegaPrime/MegaCam, a joint project of CFHT and CEA/DAPNIA, at the CFHT
which is operated by the NRC of Canada, the Institut National des Science
de l'Univers of the CNRS of France, and the University of Hawaii; {\it
  GALEX}, a
NASA funded Small Explorer Mission; SDSS and SDSS-II, which are funded by
the Alfred P. Sloan Foundation, the Participating Institutions, the
NSF, the US Department of Energy, NASA, the Japanese Monbukagakusho, the
Max Planck Society, and the Higher Education Funding Council for England;
the {\it Spitzer Space Telescope}, which is operated by the Jet Propulsion
Laboratory, California Institute of Technology under a contract with NASA;
the VLA, a facility of the NRAO, itself a facility of the NSF that is
operated by Associated Universities, Inc.; {\it XMM-Newton}, an ESA
science mission with instruments and contributions directly funded by ESA
Member States and the USA (NASA); and the W.M. Keck Observatory.  We wish to recognize and
acknowledge the very significant cultural role and reverence that the
summit of Mauna Kea has always had within the indigenous Hawaiian
community.  We are most fortunate to have the opportunity to conduct
observations from this mountain.

\bibliographystyle{apj}
\bibliography{comerford_dualAGN}

\end{document}